# Generation of Kerr combs centered at 4.5 μm in crystalline microresonators pumped by quantum-cascade lasers

**ANATOLIY A. SAVCHENKOV,[1] VLADIMIR S. ILCHENKO,[1] FABIO DI TEODORO,[2,*] PAUL M. BELDEN,[2] WILLIAM T. LOTSHAW,[2] ANDREY B. MATSKO,[1,**] AND LUTE MALEKI.[1]**

[1]OEwaves Inc., 465 N. Halstead St., Suite 140, Pasadena CA 91107
[2]The Aerospace Corp., 2310 E. El Segundo Blvd., El Segundo CA 90245
*Present address: Raytheon SAS, 2000 El Segundo Blvd. El Segundo, CA 90245
*Corresponding author: andrey.matsko@oewaves.com

**We report on the generation of mid-infrared Kerr frequency combs in high-finesse $CaF_2$ and $MgF_2$ whispering-gallery mode resonators pumped with continuous wave room temperature quantum cascade lasers. The combs were centered at 4.5 μm, the longest wavelength to date. A frequency comb wider than a half of an octave was demonstrated when approximately 20mW of pump power was coupled to an $MgF_2$ resonator characterized with quality factor exceeding $10^8$. © 2015 Optical Society of America**

http://dx.doi.org/10.1364/OL.99.099999

Optical frequency combs represent an invaluable tool for atomic and molecular spectroscopy by providing the means for precise spectral calibration and parallel interrogation of multiple absorption features over a wide range of wavelengths [1-4]. Comb generation in the mid-infrared (mid-IR) is especially attractive as it permits access to the molecular "fingerprint" region of the optical spectrum (~3-20μm wavelength), which contains strong ro-vibrational absorption features for many chemical species of environmental, industrial, medical, and military interest [5, 6].

Direct generation of frequency combs is well established for the near-IR region of the spectrum owing to the availability of mature mode-locked sources such as titanium:sapphire and ytterbium-, erbium-, and thulium-doped fiber lasers [7-9]. As mode-locked laser technology is far less advanced at longer wavelengths [10], mid-IR frequency combs have often been obtained via parametric down-conversion of near-IR lasers in nonlinear crystals, which typically results in complex, laboratory-bound optical systems [11-14]. For field applications and insertion into portable instruments, more compact and rugged architectures are desirable.

In principle, quantum cascade lasers (QCLs) would represent ideal sources for such architectures owing to their broad spectral coverage in the mid-IR (>3.5μm), excellent beam quality, room-temperature operation, high electro-optic efficiency, and inherent support for "on-chip" integration [15,16]. Recently, a specially designed, ~7μm wavelength QCL was mode-locked to generate a ~1μm-wide comb [17,18]. However, standard commercially available QCLs are generally unsuitable for direct comb generation because they exhibit relatively narrow gain bandwidths. They also are inherently difficult to mode-lock because the lifetime of their emitting states is shorter than the laser cavity roundtrip time [19].

An alternative approach to generating broadband mid-IR combs within a practical and miniaturizable platform is to use a continuous-wave (CW) QCL to optically pump a whispering-gallery-mode (WGM) microresonator [20-23]. In these devices, frequency combs (also referred to as Kerr combs) can be generated via cascaded four-wave mixing (FWM) processes initiated by frequency locking an external, "pump" laser source to a resonator WGM. Such cascade FWM processes, also known as "hyper-parametric oscillation", are often driven by modulation instability [24] and result in the emission of a spectrally broad set of sidebands spaced by one or multiple free-spectral ranges of the resonator [25]. Sidebands can be efficiently produced, even for mW-power CW pump sources, owing to the very tight optical confinement and very high quality factor (Q) of the WGM microresonator, which greatly enhance the photon density within the resonator modes, leading to strong nonlinear light/matter interactions (the optical Kerr effect). Generation of short optical pulses directly from a resonator pumped with a CW light source becomes feasible [26].

To date, nearly all realizations of Kerr combs involved near-IR pump lasers (*e.g.* ~ 1 or 1.5 μm diode lasers) and did not extend beyond ~2.5μm [23,27]. Very recently, a Kerr comb generated in a silicon micro-ring was reported to reach ~ 3.5 μm [28]. However, attaining even longer wavelengths would be highly beneficial. For example, light in the 4-5μm wavelength range exhibits good transmission through the atmosphere and overlaps with strong absorption bands of important greenhouse gases such as carbon dioxide (~4.2μm) and nitrous oxide (~4.4μm), as well as with other important species such as carbon monoxide (~4.6μm) and carbonyl sulfide (~4.8μm).

In this work we demonstrate the feasibility of combining QCLs and WGM microresonators to produce mid-IR Kerr combs within compactly integrated platforms. As our pump QCLs operated at 4.5μm, the Kerr combs were generated at the longest wavelengths observed to date.

We selected magnesium and calcium fluoride ($MgF_2$ and $CaF_2$) as the host materials for the fabrication of crystalline microresonators, mainly because of their excellent optical transmission in the mid-IR, which affords very high Q values, and anomalous group-velocity dispersion (GVD) at 4.5 μm. The anomalous GVD helps reduce the threshold power for the onset of hyper-parametric oscillation and effectively favors FWM and comb generation over parasitic nonlinearities such as Raman scattering [21, 29, 30].

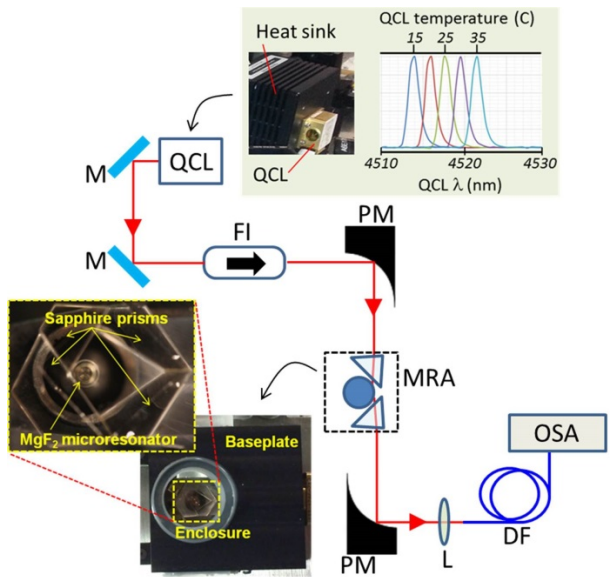

Fig. 1. Experimental setup for comb generation in the $MgF_2$ microresonator. QCL: Distributed-feedback, ~4.5μm-wavelength quantum cascade laser (inset: photograph of the laser enclosure and temperature-tuning characteristic of the laser spectral emission; M: Beam-steering mirror; FI: Faraday isolator; PM: Off-axis parabolic mirror; MRA: Microresonator assembly (inset: top-view photograph of the resonator and its sapphire-prism enclosure); L: Lens; DF: Single-mode delivery fiber; OSA: Optical spectrum analyzer.

In our first experiment, we used a commercially available, monolithically packaged distributed-feedback (DFB) QCL (Adtech Optics) emitting a single-frequency, near diffraction-limited ($M^2 \sim 1.2$) Gaussian beam with a 4.5μm central-wavelength and up to ~60mW CW power to optically pump an $MgF_2$ crystalline microresonator (see Fig. 1). The microresonator was fabricated by first cutting a $MgF_2$ cylindrical preform of ~3.6mm diameter into a thin disk, then mechanically fine-polishing its circumferential edge [31]. The resulting resonator morphology is that of a truncated oblate spheroid of 1.8 and 0.25 mm semi-axis dimensions and approximately 0.5 mm thickness [30]. The resonator GVD value can be conveniently characterized by the dimensionless parameter D, defined as [24, 25]

$$D = \gamma_i^{-1}(\nu_{i-1} + \nu_{i+1} - 2\nu_i) = -2\pi c \beta_2(\nu_i) \nu_{FSR}^2 n_i^{-1} \gamma_i^{-1} \quad (1)$$

Here, "i" is a mode number, $\nu_i$ ($\gamma_i$) is the optical frequency (half width at half-maximum) of the $i^{th}$ resonator WGM, $n_i$ is the refractive index of the host material corresponding to the frequency $\nu_i$, c is the speed of light in vacuum, and $\nu_{FSR} = (\nu_{i-1} + \nu_{i+1})/2$ is the free spectral range (FSR) of the resonator, and $\beta_2$ is the GVD of the resonator host material.

Based on the known material dispersion and resonator morphology of and assuming $\gamma_i \sim 170$ kHz, we obtain $\nu_{FSR} \sim 19.3$ GHz, hence D ~ 2 for $MgF_2$ resonator. The estimated D value is very suitable to the generation of stable, mode-locked Kerr combs and is known to enable the observation of a diverse variety of comb structures including hyper-parametric oscillations, type-II combs, and Turing patterns, which can all be accessed by judiciously adjusting the detuning between the pump laser optical frequency and WGM spectrum [24, 33-35]. Smaller values of D can result in broader comb generation for a given pump power and resonator Q, although the number of possible comb regimes would increase, which may degrade stability [26].

The resonator Q factor exceeded $10^{10}$ at ~ 1550 nm, but was reduced to ~$2\times10^8$, (corresponding to ~330 kHz bandwidth) at ~4.5 μm, due to greater background optical absorption related to multi-phonon effects and residual organic impurities [36]. The microresonator was mounted within an optically transparent enclosure to protect it from dust and humidity, and allow transportability and operation in non-clean-room environments (see Fig. 1). In our design, the enclosure walls also serve as a pair of evanescent-field coupling prisms to direct light in and out of the resonator [37]. Uncoated sapphire was chosen as the enclosure material by virtue of its environmental stability, excellent optical transmission at mid-IR wavelengths, and refractive index (~1.65 at ~4.5μm), which is high enough to permit optical coupling into $MgF_2$ (n ~ 1.34), yet low enough to keep Fresnel reflections down to ~6%. The microresonator protective enclosure was installed on a thermalized heat sink, which provided temperature control to within $10^{-3}$ K. The temperature control and judicious use of materials having dissimilar thermal expansion coefficients for resonator, coupling prism, and mounting substrate, enabled the fine adjustment of the resonator/coupling-prism air gap dimension to within ~100nm.

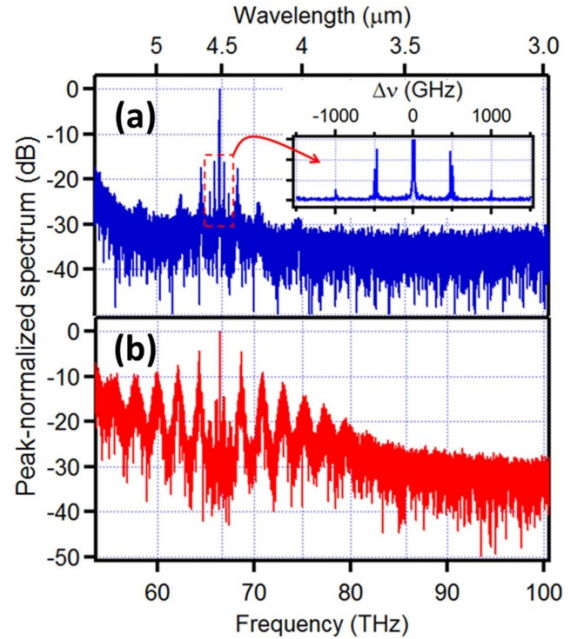

Fig. 2. Kerr frequency combs in the $MgF_2$ microresonator. Logarithmic-scale, peak-normalized spectra of Kerr combs, obtained by optically pumping the $MgF_2$ microresonator with ~55mW power at ~4.5 mm wavelength. The spectra in (a) and (b) were obtained for different values of detuning between the pump laser frequency and resonator modes. A narrow-spaced (~500 GHz) comb, spectrally localized in the vicinity of the pump, persisted for a relatively wide range of detunings. A linear-scale spectrum of this comb plotted vs. frequency difference from the pump is shown in the inset of (a). A more widely spaced (~2.1 THz) comb, shown in (b), could be obtained for a narrower range of detunings and spanned half an octave.

The collimated QCL output beam was transmitted through a mid-IR Faraday isolator (Innovation Photonics), then formatted and injected into the microresonator enclosure by a gold-coated off-axis parabolic mirror having 50.8mm reflected focal length. A similar parabolic mirror was used to collect the beam exiting the microresonator. As the optical isolator blocks back reflections from the microresonator, thermal frequency locking, which is made possible by the positive thermo-optic (d$n$/dT) coefficient of $MgF_2$, was used to passively couple the QCL to resonator WGMs [38]. To achieve thermal locking, we varied the driving QCL voltage, thus finely sweeping its output frequency around resonator resonances.

Kerr combs could be generated for QCL output powers as low as ~2mW, when the prism/resonator gap was adjusted to yield a loaded resonator Q factor of ~ 1.5 × $10^8$ (~ 450 kHz bandwidth). We observed two distinct combs characterized by mode spacing of ~ 2100 and ~500 GHz (Fig. 2), respectively, which could be generated at smaller and larger offsets of the QCL laser frequency relative to the WGM mode frequency. As the QCL power was increased to ~55 mW, the wider-spaced Kerr comb stretched to more than half of an octave (~3.7 to ~5.5 μm, corresponding to over 25 THz), while the narrow-spaced comb remained confined to a ~ 2.5 THz region around the pump wavelength (Fig. 2b).

To illustrate the potential for compact and robust packaging of QCL-pumped WGM microresonators, we designed and built a demonstrator device consisting of a 4.5μm-wavelength DFB QCL chip (Adtech Optics, up to ~50mW emitted power, near-diffraction-limited Gaussian beam quality), anti-reflection coated silicon collimating lenses, barium fluoride ($BaF_2$) prism coupler, and truncated-spheroid microresonator (3 and 0.21 mm semi-axis dimensions, ~0.5mm thickness), all mounted on a thermo-controller base (see Fig. 3). For this device, the microresonator host material was calcium fluoride ($CaF_2$), which shares many relevant optical and mechanical properties of $MgF_2$. In particular, the D value for this resonator (obtained by applying Eq. 1) is ~ 0.35, which is also viable for stable Kerr comb generation at low threshold power. The device concept and overall dimensions (approximately 75×50×10 mm) are similar to those of previously reported units, in which we integrated WGM microresonators and ~1.5μm-wavelength telecom-type DFB diode lasers [39]. The insertion loss of the combined resonator/prism/coupling-lens subsystem was measured to be < 3dB. Because the negative thermo-optic coefficient of $CaF_2$ makes thermal locking impractical, the QCL frequency was self-injection locked to resonator WGMs by letting light generated via Rayleigh scattering within the microresonator back into the QCL (no optical isolation is introduced between QCL and microresonator for this purpose) [39].

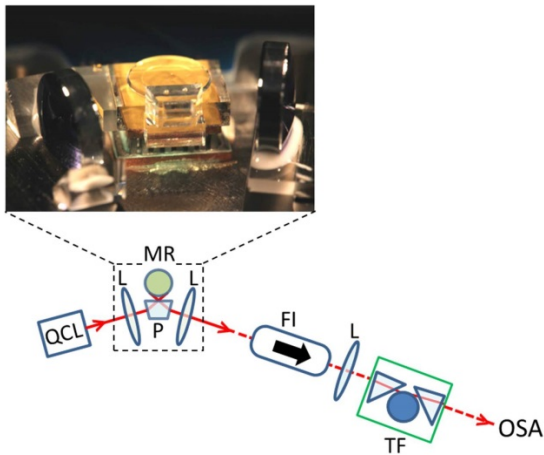

Fig. 3. Experimental setup for comb generation and measurement in the $CaF_2$ microresonator. QCL: Distributed-feedback, ~4.5mm-wavelength quantum cascade laser; L: Lens; P: Evanescent-field coupling prism, MR: Microresonator (inset: photograph of the compactly packaged resonator/coupling-prism/coupling-lens assembly); FI: Faraday isolator; TF: Tunable Lorentzian filter consisting of a second, $MgF_2$ microresonator equipped with evanescent-field coupling prisms and installed on temperature-controlled base; OSA: Optical spectrum analyzer.

In the $CaF_2$ device, we observed the generation of a Kerr comb at pump power as low as ~ 15 mW. As we maximized the QCL power within the resonator to ~30mW, the comb stretched to a width of approximately 10 THz (Fig. 4). Similar to the case of the $MgF_2$ microresonator, we could obtain different Kerr comb line spacings for different values of the QCL frequency to WGM offset. Moreover, for both resonators, the amplitude profiles of the comb envelopes, as well as the overall process of comb formation, appeared very similar to Kerr comb generation regimes driven by the modulation instability usually observed and analyzed in fused silica microresonators pumped at ~1.5μm [40].

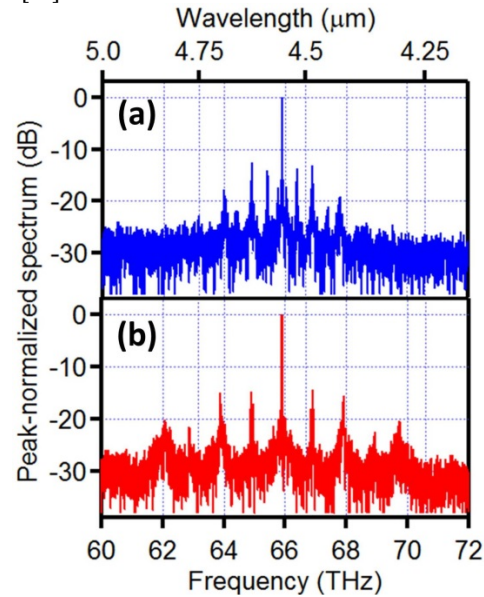

Fig. 4. Kerr frequency combs in the $CaF_2$ microresonator. (a) Logarithmic-scale, peak-normalized spectrum of Kerr comb obtained by optically pumping the $CaF_2$ microresonator with ~30mW power emitted by a ~4.5mm wavelength, distributed-feedback quantum cascade laser (QCL) self-injection-locked to a resonator WGM. (b) Kerr comb spectrum obtained for the same pump power, but with different QCL/resonator-mode frequency detuning.

In both experiments, the spectral data was obtained by directing the resonator output beam into a Fourier-transform optical spectrum analyzer (OSA) (Thorlabs OSA205) equipped with single-mode, mid-IR transmissive input optical fiber (IRflex). The output power was measured with a mercury-cadmium telluride (HgCdTe) photodetector (Thorlabs PDA10JT). Since the OSA spectral resolution (8 GHz at ~ 4.5 μm) was marginal for measuring the spectral linewidth of individual comb lines and fully discriminating dense combs, we fabricated a third, stand-alone WGM microresonator (host material: $MgF_2$) of similar design to those described above.

By adjusting the spatial separation between the resonator surface and evanescent-field prism coupler, we could vary the resonator bandwidth in the 0.3 to 300 MHz range, thus effectively obtaining a tunable-bandwidth, first-order Lorentzian band-pass filter. By transmitting the $CaF_2$ resonator output beam through this $MgF_2$ filter prior to directing it to the OSA, we were able to fit individual comb lines (recorded over an ~ 1ms-long integration time) to Lorentzian profiles, which yielded a half-width at half maximum of ~60kHz.

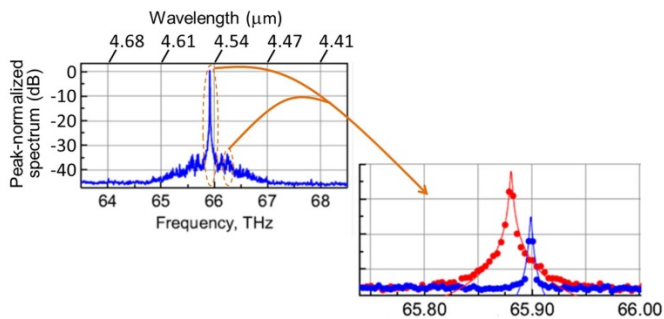

Fig. 5. Measurement of comb linewidth. Lorentzian fit (solid lines) to the pump (red dots) and to a Kerr comb spectral (blue dots) lines observed at the exit of the $CaF_2$ microresonator optically pumped by ~20mW-power, 4.5mm light. The two lines were isolated from the full spectrum by using a separate $MgF_2$ microresonator as a tunable ~4MHz full-width-at-half-maximum (FWHM) band-pass filter. The Lorentzian fit yielded FWHM ~ 60kHz for both lines.

It is interesting to note that while both $CaF_2$ and $MgF_2$ resonators are characterized with anomalous GVD, we observed Turing pattern-like frequency combs [34] in the magnesium fluoride one (Fig. 2). The calcium fluoride resonator led to demonstration of both Turing patterns-like (Fig. 4) and multi soliton-like (Fig.5) comb envelopes. The later ones were spectrally narrow, which corresponds to the large GVD of the resonators. In depth study of the regimes was hindered by low resolution of our optical spectrum analyzer as well as unavailability of a high speed photodiode.

In conclusion, we have demonstrated the generation of the longest-wavelength Kerr combs to date by optical pumping of crystalline WGM microresonators with 4.5μm-wavelegnth DFB QCLs. The combs were observed for both thermal and self-injection passive locking of the QCL frequency to that of the resonator modes, and the comb width could exceed half an octave. Our experiments show that the generation of long-wavelength mid-IR combs can be obtained by combining QCLs and fluoride microresonators within integrated platforms in a manner similar to well-established near-IR systems that are based, for example, on telecom-type diode lasers and silica resonators. We stress that, owing to the maturity and availability of QCLs operating at even longer wavelength, this approach naturally offers the unique and attractive possibility of extending Kerr comb generation into the far infrared and THz domain.

The authors acknowledge support from Air Force Office of Scientific Research under contract No. FA9550-12-C-0068. A.M. also acknowledges illuminating discussions with Dr. T. Kippenberg. After this work was completed the authors became aware of similar study performed at EPFL [41].